\documentclass{vgtc}                          

\ifpdf
  \pdfoutput=1\relax                   
  \usepackage{graphicx}                
  \DeclareGraphicsExtensions{.pdf,.png,.jpg,.jpeg} 
\else
  \usepackage{graphicx}                
  \DeclareGraphicsExtensions{.eps}     
\fi%

\graphicspath{{figures/}{pictures/}{images/}{./}} 

\usepackage{microtype}                 
\PassOptionsToPackage{warn}{textcomp}  
\usepackage{textcomp}                  
\usepackage{mathptmx}                  
\usepackage{times}                     
\usepackage{cite}                      
\usepackage{tabu}                      
\usepackage{booktabs}                  

\usepackage{amsmath,amsfonts}
\usepackage{array}
\usepackage[caption=false,font=normalsize,labelfont=sf,textfont=sf]{subfig}
\usepackage{stfloats}
\usepackage{url}
\usepackage{verbatim}
\usepackage{colortbl}
\usepackage{makecell}
\usepackage{wrapfig}
\usepackage{tabularx}

\usepackage{enumitem}

\usepackage{algorithm}
\usepackage{algpseudocode}

\onlineid{0}

\vgtccategory{Research}

\vgtcinsertpkg

\teaser{
    \centering
    \includegraphics[width=\textwidth]{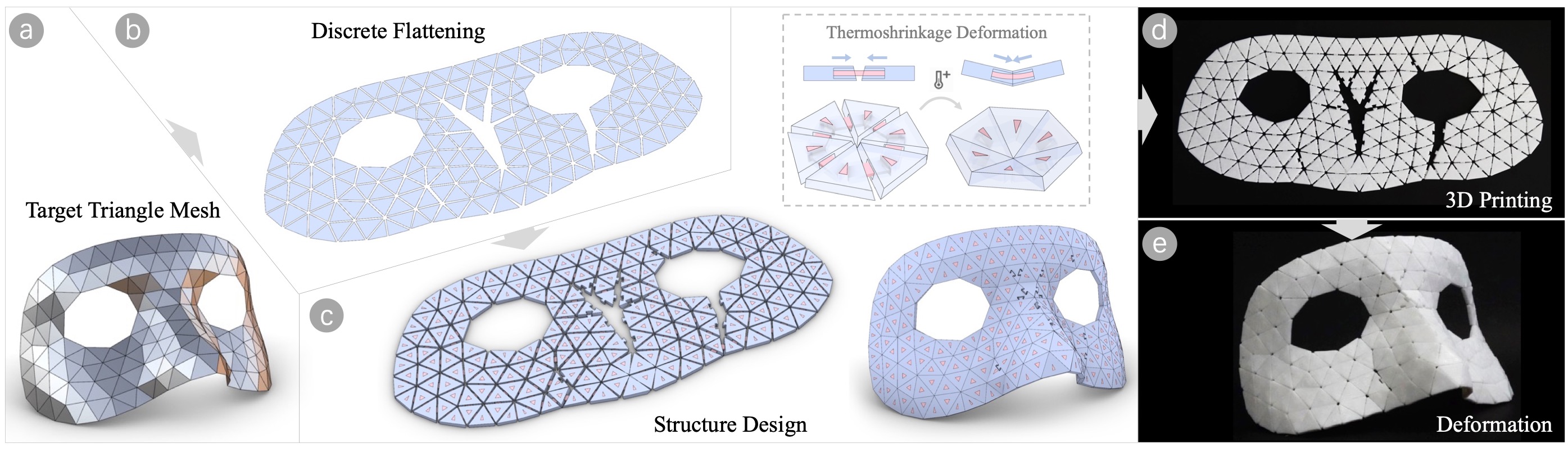}
    \caption{The pipeline of FreeShell. Starting from a triangle mesh as the target surface (a), our method automatically produces a suitable 2D layout for all triangles as a 2D mesh (b).A flat plate structure is constructed based on the 2D mesh (c). Finally, the flat plate is fabricated by 3D printing (d) and the result is formed that approximates the target mesh by thermoshrinkage deformation (e).}
    \label{topfig}
}

\title{FreeShell: A Context-Free 4D Printing Technique for Fabricating Complex 3D Triangle Mesh Shells}

\author{Chao Yuan %
\and Nan Cao\thanks{Corresponding author. e-mail: nan.cao@gmail.com} %
\and Xuejiao Ma %
\and Shengqi Dang}
\affiliation{\scriptsize Intelligent Big Data Visualization Lab, Tongji University}

\abstract{Freeform thin-shell surfaces are critical in various fields, but their fabrication is complex and costly. Traditional methods are wasteful and require custom molds, while 3D printing needs extensive support structures and post-processing. Thermoshrinkage actuated 4D printing is an effective method through flat structures fabricating 3D shell. However, existing research faces issues related to precise deformation and limited robustness. Addressing these issues is challenging due to three key factors: (1) Difficulty in finding a universal method to control deformation across different materials; (2) Variability in deformation influenced by factors such as printing speed, layer thickness, and heating temperature; (3) Environmental factors affecting the deformation process. To overcome these challenges, we introduce FreeShell, a robust 4D printing technique that uses thermoshrinkage to create precise 3D shells. This method prints triangular tiles connected by shrinkable connectors from a single material. Upon heating, the connectors shrink, moving the tiles to form the desired 3D shape, simplifying fabrication and reducing material and environment dependency. An optimized algorithm for flattening 3D meshes ensures precision in printing. FreeShell demonstrates its effectiveness through various examples and experiments, showcasing accuracy, robustness, and strength, representing advancement in fabricating complex freeform surfaces.
} 

\CCScatlist{ 
  \CCScat{}{Computing methodologies}{}{Shape modeling};
  \CCScat{}{Applied computing}{}{Computer-aided design}
}

\begin{document}
\maketitle

\section{Introduction}

Freeform thin-shell surfaces are extensively utilized across various domains, including lightweight construction, ergonomic wearable devices, ergonomic electronic products, and medical instruments including scar pressure masks and orthopedic shell products that require customization to fit human curvature. These shells offer significant value in enhancing comfort, aesthetics, and functionality. However, their fabrication involves more complex and expensive processes compared to planar or single-curved surfaces, especially for personalized manufacturing. Subtractive manufacturing methods such as milling result in material waste, while thermoforming and casting require custom molds. Additionally, 3D printing often requires extensive support structures and complex post-processing.

One approach to addressing these issues is to use 3D printing techniques with self-shaping materials. These materials printed as flat forms can be transformed into specific geometries under certain stimuli such as temperatures, humidity, and light. Therefore, the process is commonly referred to as 4D printing, which has shown significant potential in reducing fabrication costs, facilitating transportation, and enabling on-site deployment. Fused Deposition Modeling (FDM) is one of the most affordable and widely used 3D printing techniques. It often utilizes the thermoshrinkage of materials like polylactic acid (PLA) to achieve 4D printing, where deformation occurs due to the materials shrinking when heated. Existing studies have shown that designing specific printing paths and structures can create localized variations in thermoshrinkage within materials. This enables control over the direction and extent of material bending, resulting in the desired deformation when heated. However, achieving precise deformation remains a challenge. Current research faces issues related to limited robustness, which significantly hinders the widespread adoption of this technique.

Addressing these issues is usually difficult due to three critical challenges: (1) It is difficult to find a universal printing method to control the deformation for different materials precisely; (2) Even for the same material, deformation is influenced by various factors such as printing speed, layer thickness, and heating temperature. Controlling these factors precisely is difficult; (3) Even if the above conditions are effectively controlled, the printing environment can still affect the deformation process. Precisely controlling the environment is a challenge. 

To address these challenges, we introduce FreeShell, a 4D printing technique that is robust enough for accurately fabricating complex 3D shells regardless of material properties, printing conditions, and the printing environment, i.e., fully context-free. We achieve this goal by printing a network of scattered triangular tiles connected by shrinkable connectors. Both the connectors and tiles are printed using the same material but in different ways, resulting in different shrinking rates. Once heated, the connectors shrink, causing the tiles to move. The edges of adjacent tiles are tangent to each other at an angle along the normal direction of the target surface (Fig.~\ref{topfig}c), thus ensuring accuracy. The contributions of this paper are:

\begin{itemize}[leftmargin=*]
    \item \textbf{Innovative 4D Printing Technique:} We introduce a novel context-free 4D printing technique that utilizes thermoshrinkage actuated curvature to fabricate 3D shells. This method requires only a single type of material and eliminates the need for support structures or segmentation, simplifying the fabrication process. The method can produce accurate 3D shells without requiring precise control over the shrinkage rate of the printing material, significantly reducing dependence on the material properties, printing conditions, and environment.
    
    \item \textbf{Optimized 3D Mesh Flattening:} We introduce an optimized method for flattening and arranging the tiles on a 2D plane within the triangle mesh of the target shell. The algorithm strategically cuts the mesh to minimize distortions and maintain precision during the flattening process. 

    \item \textbf{Demonstration and Evaluation:} We demonstrated the power of our technique through a number of example cases and via a series of experiments including printing performance, accuracy, robustness, and strength.
\end{itemize}

\section{Related Work} \label{sec2}
Thermoshrinkage actuated 4D printing uses the thermoshrinkage properties of materials to transform flat printed forms into specific geometric shapes. This process primarily involves studying self-shaping techniques and flattening techniques for target surfaces.

\subsection{Self-Shaping Techniques}
\paragraph{Materials of Self-Shaping Surfaces}
Self-shaping surfaces refer to a category of intelligent materials or structures capable of changing their shape in response to external stimuli such as temperature, humidity, light, electric fields, or magnetic fields. The realization of self-shaping surfaces relies on special structural designs and internal stress distributions within the materials. Various materials are used to create self-shaping surfaces, including hydrogels~\cite{kim_designing_2012, na_grayscale_2016, sydney_gladman_biomimetic_2016}, liquid crystal elastomers~\cite{aharoni_universal_2018}, hygroscopic wood materials~\cite{gronquist_analysis_2019, cheng_programming_2021, luo_autonomous_2023}, stretchable membrane materials~\cite{guseinov_curveups_2017, perez_computational_2017, guseinov_programming_2020, jourdan_computational_2022}, and deformable pasta during cooking~\cite{wang_transformative_2017, tao_morphlour_2019}.

\paragraph{Thermoshrinkage Actuated 4D Printing}
Thermoshrinkage actuated 4D printing holds significant research potential due to its advantages of ease of fabrication and simple triggering conditions. Existing studies have utilized the shape memory effect of materials like PLA to achieve 4D printing. The basic principle is that after extrusion and rapid solidification, materials maintain internal tensions along their deposition trajectories. When these materials are reheated to their glass transition temperature, they shrink along the directions of these trajectories. Therefore, researchers utilize these properties to design printing paths. Some studies directly exploit shrinkage for deformation~\cite{wang_4dmesh_2018, sun_shrincage_2021, chalissery_highly_2022, moon_shrinkcells_2022, wang_4dcurve_2023}, while others employ multi-layered printing structures to achieve specific deformations. For instance, deformation can be achieved by adjusting the thickness of different layers~\cite{gu_geodesy_2019}, bending can be induced by the perpendicular permutation of bilayer paths~\cite{van_manen_programming_2017, yu_material_2020, tao_4doodle_2023}, twisting can be realized by the permutation of bilayer paths at specific angles~\cite{wang_morphingcircuit_2020}, and more flexible bending can be attained through customized path arrangements~\cite{wang_-line_2019}. Additionally, deformation is accomplished by using multi-materials with different shrinkage rates in different layers, such as shrinkable PLA and non-shrinkable TPU materials~\cite{an_thermorph_2018, deshpande_fab4d_2022}. These studies focus on designing various tool paths and controlling related printing parameters like speed, layer thickness, and printing temperature to construct the design space for material deformation. However, these studies face challenges in accurately approximating free-form surfaces, focusing more on exploring applications based on their deformation capabilities.

\subsection{3D Flattening Technique}
\paragraph{Surface Flattening}
Surface flattening is a widely researched topic in computer graphics, focusing on mapping non-developable surfaces from the 3D domain to the 2D domain with minimal distortion, which can simplify manufacturing and reduce costs. For example, methods such as origami approximation of target surfaces~\cite{tachi_origamizing_2010, jiang_freeform_2020}, stripification~\cite{mitani_making_2004, schuller_shape_2018}, and developability optimization~\cite{stein_developability_2018, ion_shape_2020, binninger_developable_2021, zhao_developability-driven_2022} are used to flatten target surfaces. However, these methods do not involve materials with stretchable deformation, often resulting in segmentation to reduce distortion, which complicates assembly.

\paragraph{Deployable Structures}
Therefore, some studies employ deformable materials to approximate target surfaces, such as auxetic structures~\cite{konakovic_beyond_2016, konakovic-lukovic_rapid_2018, chen_bistable_2021}, elastic structures~\cite{malomo_flexmaps_2018}, and inflatable structures~\cite{siefert_programming_2020, panetta_computational_2021}. Relevant research focuses on controlling the material's local anisotropic shrinkage, such as elastic fabrics~\cite{guseinov_curveups_2017, perez_computational_2017, guseinov_programming_2020}. These methods use pre-stretched elastic fabrics to self-deform flat structures into target shapes. However, using elastic fabrics involves a more tedious process of fixing them to the printing plate. Our method is conceptually similar to "CurveUps"~\cite{guseinov_curveups_2017}, utilizing tiles to achieve deployment. However, "CurveUps" employs elastic membranes, resulting in non-rigid structures. The nature of the pre-stretched membrane causes the shrinkage force to decrease closer to the anchor points, which may lead to insufficient shrinkage at the structural boundary. Additionally, the fabrication process is complex due to the necessity of removing support structures and adhering elastic membranes to tiles. In contrast, our approach adopts a discrete structure in which thermoshrinkable connectors generate sufficient shrinkage force at each position, ultimately forming a rigid structure. Our method utilizes only a single material and eliminates the need for support
structures, simplifying the fabrication process.

\paragraph{Thermoshrinkage-based Flattening}
In the field of 4D printing, research focuses on developing flattening methods based on the thermoshrinkage of materials. The key challenge is correlating the local distortions of the flattened surface with the local shrinkage rate of the material. One approach involves transforming the target surface into a locally shrinkable wireframe. For instance, Wang et al.~\cite{wang_4dmesh_2018} preprocess the target surface into a wireframe and then flatten it using conformal mapping, where the distortion of the wireframe's edges corresponds to the shrinkage rate. Boley et al.~\cite{boley_shape-shifting_2019} utilize the different thermoshrinkages of two materials to control localized deformation of the wireframe. However, these wireframe-based methods face challenges in precision of approximation. Another approach employs multi-layer printing to control the shrinkage rate. For instance, Gu et al.~\cite{gu_inverse_2020} convert the target surface into a height field to form corresponding continuous printing paths, where the shrinkage rate is controlled by continuous multi-layer paths with varying thicknesses. However, this method is only applicable to surfaces without self-intersections in their 2D projection. Jourdan et al.~\cite{jourdan_shrink_2023} utilize different printing path directions in each layer to induce anisotropic shrinkage, thereby deforming flat structures into target surfaces, which can be used for more freeform surfaces. However, these methods require precise control over the material's shrinkage rate at different locations, leading to the need for stringent control of the printing conditions and the printing environment. Additionally, since these methods rely on multi-layer printing to control shrinkage rates, they cannot arbitrarily change the shell's thicknesses to meet specific design needs, such as increasing thickness for material strength, which limits their applicability.

\section{Method Overview} \label{sec3}

\begin{figure}[ht]
    \centering
    \includegraphics[width=\linewidth]{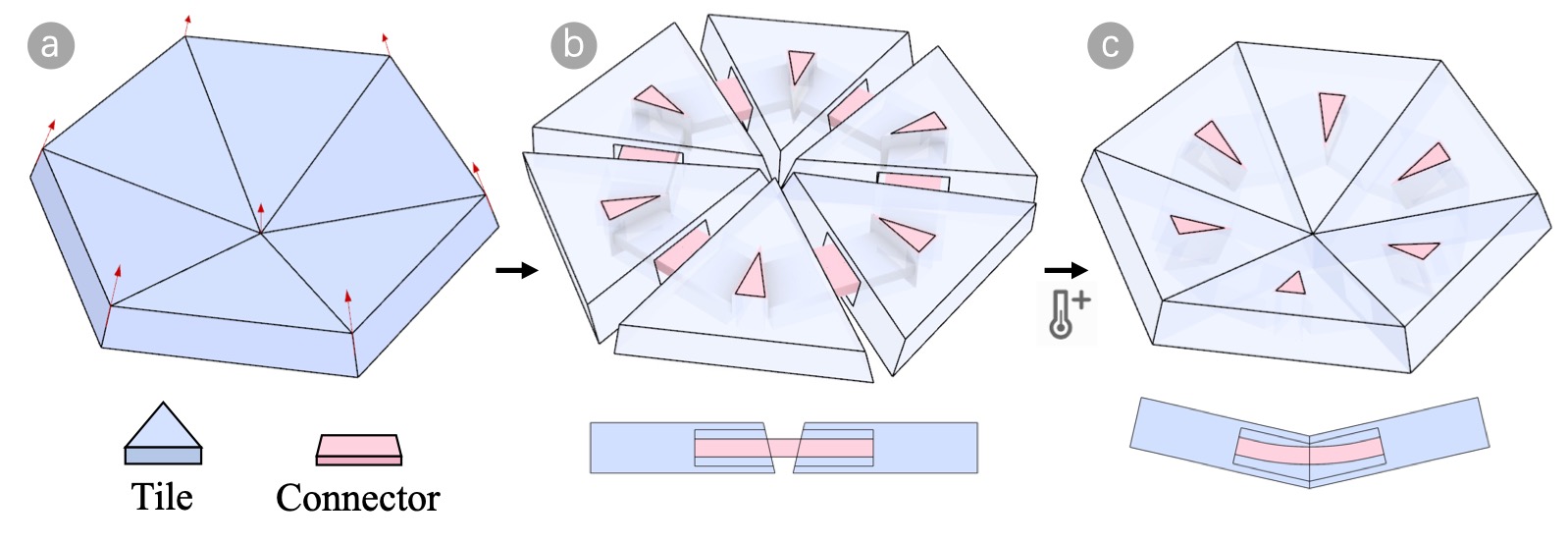}
    \caption{Basic idea of thermoshrinkage actuated curvature. The target shell consists of tiles that contain the curvature of the target shell (a). The flat plate consists of tiles and connectors (b). The connectors drive the tiles to form the target shell (c).}
    \label{basis}
\end{figure}

\paragraph{Goals} Our goal is to design a tiles-based flat plate structure that can transform into the target shell through thermoshrinkage actuated curvature. The structure consists of rigid tiles shaped like frustums, designed to match the curvature of the target shell. These tiles configure themselves into the target shell, as long as all thickened edges between adjacent tiles make contact. The tiles are connected by shrinkable connectors, which harness the shape memory effect to store tension and provide the driving force for self-deformation. Upon heating, these connectors shrink and apply forces to pull the adjacent tiles together, thus achieving the desired shell shape (see Fig.~\ref{basis}). Our method does not require precise control over the printing conditions and environment, as long as a certain shrinkage threshold is met to ensure that adjacent tiles are tightly contacted. Therefore, we focus on the flat plate design.

\paragraph{Methodology} Finding the optimal design for such a flat plate is challenging. It is crucial to achieve an appropriate 2D layout of the tiles so that adjacent tiles make tight contact within the limited shrinkage of the connectors. To address this problem efficiently, we propose a \textit{discrete flattening} method to generate the 2D layout. First, we uniformly remesh the input 3D surface into a triangular mesh, with the edge length of the triangles suitable for 3D printing. Then, we compute a suitable 2D mesh using the \textit{discrete flattening} method, ensuring all triangles in the mesh are rigid and non-intersecting, with gaps between adjacent triangles less than the shrinkage length of the connectors. The \textit{discrete flattening} method provides the initial shape of the flat plate and arranges these tiles into a 2D layout. Furthermore, based on the 2D layout, connectors are constructed to join adjacent tiles, forming a flat plate for 3D printing.

\paragraph{Workflow}
Figure \ref{topfig} illustrates the workflow. First, our process begins with the target triangular mesh as input. The target 3D mesh is then flattened into a 2D mesh using our \textit{discrete flattening} method (Section \ref{sec_flatten}). Subsequently, all the tiles and connectors are constructed on the 2D mesh to form the flat plate (Section \ref{sec_structure}). Finally, the target shell is fabricated through 3D printing and thermoshrinkage actuated deformation (Section \ref{sec_implement}).

\section{Model} \label{sec4}

In this section, we formally introduce our model and design parameters. We begin with the geometric model, followed by the requirements for the 2D layout and the definition of a comprehensive energy function. Finally, we present the key ideas to solve the optimization problem.
\begin{figure}[ht]
    \centering\includegraphics[width=0.5\linewidth]{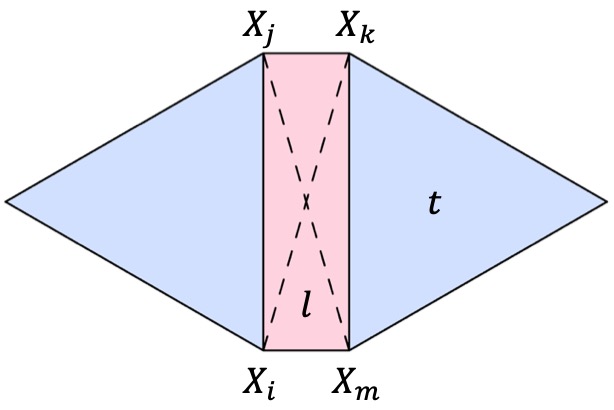}
    \caption{Two adjacent triangles with notation}
    \label{adj}
\end{figure}

\paragraph{Geometry}
Each tile is constructed by thickening the corresponding triangle in the target mesh. Given a target 3D mesh $\hat{M} \in \mathbb{R}^{3}$, we compute a suitable 2D mesh $M \in \mathbb{R}^{2}$ as a 2D layout to pack these tiles. The vertex set, edge set, linkage set, and triangle set in $\hat{M}$ are denoted as $\hat{X}, \hat{E}, \hat{L}, \hat{T}$, respectively. Correspondingly, these sets in $M$ are denoted as $X, E, L, T$, where $l \in L$ as a linkage connects each pair of adjacent triangles. A detailed illustration can be seen in Figure \ref{adj}. The linkage $l$ contains two vertex index sets of the edges: $im$ and $jk$. Therefore, for each pair of adjacent triangles, the gap value $\mathbf{a}$ of the adjacent triangles is denoted as:
\begin{equation}
    \mathbf{a} = (||e_{im}|| + ||e_{jk}||) / 2
\end{equation}
where $e_{im} = X_{i} - X_{m}$ denotes an edge vector of the linkage. Upon heating, the 3D-printed connectors release tension to pull the tiles together. Eventually, the tiles are contacted, producing contact forces that balance the tension.

\paragraph{Goal}
Our goal is to flatten $\hat{M} \in \mathbb{R}^{3}$ into $M \in \mathbb{R}^{2}$ to achieve a suitable 2D layout of all triangles. During deformation, all adjacent tiles can be pulled together by the tension of the connectors, ensuring that the thickened edges between adjacent tiles are aligned. Specifically, $M$ satisfies the following three requirements: (1) each triangle is rigid; (2) $\mathbf{a}$ is less than the connector's shrinkage length; and (3) adjacent triangles are aligned without shearing.

In addition, to simplify assembly process, the flat plate is non-segmented. We use a graph $\mathcal{G}: (\mathcal{N}, \mathcal{L})$ to define this requirement, where $\mathcal{N}$ is a set of triangles that are represented as nodes, and $\mathcal{L}$ is a set of linkages that are represented as edges. The graph must be connected, meaning that for any two nodes $\mathcal{N}_{i}, \mathcal{N}_{j} \in \mathcal{N}$, there exists a path $\mathcal{P} \subseteq \mathcal{L}$ that connects $\mathcal{N}_{i}$ and $\mathcal{N}_{j}$.

We formulate this goal as a constrained nonlinear optimization problem, where soft constraints represent the desired properties of the final result, while hard constraints ensure that the non-segmented assembly requirement are met. The optimization problem is formulated as follows:
\begin{equation}
\begin{aligned}
    \underset{X}{\mathrm{minimize}} \quad & E_{\mathrm{opti}} = \omega_{\mathrm{rigid}}E_{\mathrm{rigid}} + \omega_{\mathrm{gap}}E_{\mathrm{gap}} + \omega_{\mathrm{fair}}E_{\mathrm{fair}} \\
    \mathrm{subject~to} \quad &  N_{\mathcal{N}} = N_{T}
\end{aligned}
\label{eq1}
\end{equation}
where $E_{\mathrm{rigid}}$ measures the shape distortion of each triangle in $M$, $E_{\mathrm{gap}}$ represents the tightness between adjacent tiles, and $E_{\mathrm{fair}}$ is a fairness preservation term used to measure the shear between adjacent triangles. These energies are balanced by three positive weights: $\omega_{\mathrm{rigid}}$, $\omega_{\mathrm{gap}}$, and $\omega_{\mathrm{fair}}$. The quantities $N_{\mathcal{N}}$ and $N_{T}$ denote the number of nodes in $\mathcal{G}$ and the number of triangles in $M$, respectively, ensuring that the flat plate remains non-segmented.

\paragraph{Rigidity Energy} In order to minimize the deformation of tiles, we introduce a rigidity energy term to measure the shape distortion of triangles as follows:
\begin{equation}
    E_{\mathrm{rigid}} = \sum_{t \in T} \sum_{(i,j) \in t} \mathbf{\bar{E}}^{-2}{\left(||X_{i} - X_{j}|| - ||\hat{e}_{ij}|| \cdot \theta \right)^2}
    \label{eq2}
\end{equation}
where $t$ denotes an index set of vertices in a triangle, $\mathbf{\bar{E}}$ denotes an average length of all edges in $\hat{M}$ and $\theta$ is an amplification factor. $\mathbf{\bar{E}}^{-2}$ makes the energy term dimensionless.

\paragraph{Gap Energy} To ensure that adjacent tiles are pulled together by the connectors after deformation and to reduce the intersection of triangles in $M$, we introduce a gap energy term to measure the deviation between the target gap value and the edge lengths of all linkages as follows:
\begin{equation}
    E_{\mathrm{gap}} = \sum_{l\in L} \sum_{im \in l} \mathbf{\bar{E}}^{-2}\left(||e_{im}|| - \mathbf{d} \cdot \gamma \right) ^ 2
    \label{eq3}
\end{equation}
where $\mathbf{d}$ denotes the target gap value related to the shrinkage length of the connectors, and $\gamma$ is a scaling factor.

\begin{figure}[ht]
    \centering
    \includegraphics[width=\linewidth]{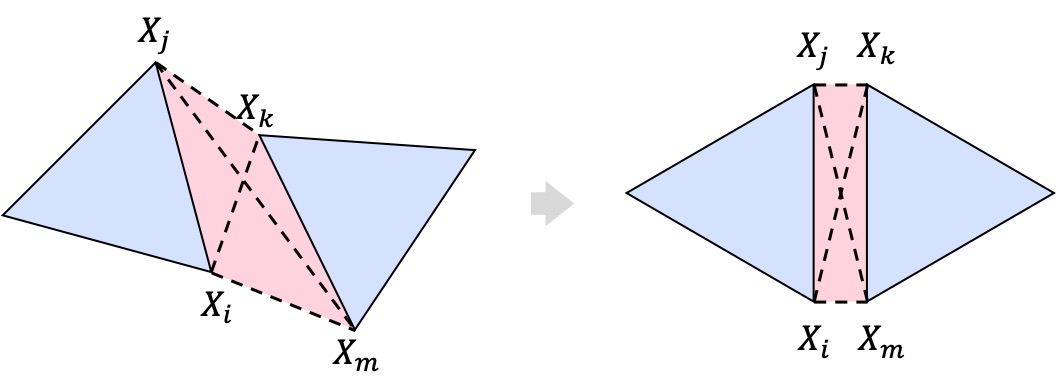}
    \caption{Shear avoidance: The fairness energy term aims to make the interior angles of the linkage as close to right angles as possible. Specifically, $||X_{i} - X_{m}||$ and $||X_{j} - X_{k}||$ both represent the short edge lengths of the linkage, which are expected to be equal to $\mathbf{d}$.}
    \vspace{-3mm}
    \label{hypotenuse}
\end{figure}

\paragraph{Fairness Energy} To ensure the alignment of the thickened faces between adjacent tiles when they come into contact, we introduce a fairness energy term. This term prevents adjacent triangles from shearing each other and penalizes misalignments between tiles:
\begin{equation}
    E_{\mathrm{fair}} = \sum_{h\in H} \sum_{(i,k) \in h} \mathbf{\bar{E}}^{-2}\left(||e_{ik}|| - \textbf{h}_{ik} \right) ^ 2
    \label{eq4}
\end{equation}
where $h$ denotes the index set of vertices in a hypotenuse, $H$ is the set of all hypotenuses, and $\textbf{h}_{ik}$ represents the target length of the hypotenuse, which satisfies the relationship as follows:
\begin{equation}
    \textbf{h}_{ij} = \sqrt{||X_{i} - X_{j}|| ^ 2 + \mathbf{d} ^ 2} = \textbf{h}_{km} = \sqrt{||X_{k} - X_{m}|| ^ 2 + \mathbf{d} ^ 2}
    \label{eq5}
\end{equation}
where the $||X_{i} - X_{j}||$ and $||X_{k} - X_{m}||$ both represent long edge lengths in the linkage. As illustrated in Figure~\ref{hypotenuse}, we utilize properties based on right triangles to avoid shearing between adjacent triangles.

\begin{figure*}[!t]
    \centering
    \includegraphics[width=\textwidth]{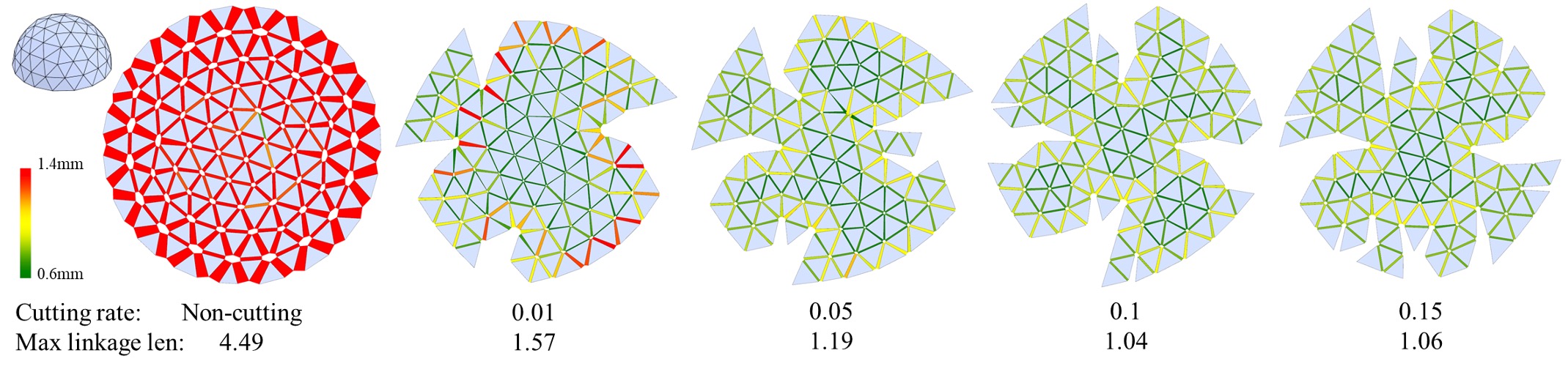}
    \caption{Results with different cutting rate. A low cutting rate results in large gap values, whereas a higher cutting rate results in smaller gap values}
    \vspace{-3mm}
    \label{cutting}
\end{figure*}

\paragraph{Challenges} Solving the optimization problem in Eq. (\ref{eq1}) exists three main challenges: First, directly flattening $\hat{M}$ into $M$ while satisfying these energy terms is a non-convex optimization problem, making it difficult to achieve a global optimal solution. Then, since the existence of local distortion when flattening non-developable surfaces and preserving rigidity in all triangles, some gaps between adjacent triangles could exceed $\mathbf{d}$. Finally, separating all mesh triangles to create a 2D layout can lead to intersection issues.

\paragraph{Key Ideas} We introduce key ideas to solve these challenges. We first compute a conformal mapping from the target mesh $\hat{M}$ to a 2D domain based on the ARAP method~\cite{sorkine_arap_2007}, which serves as a pre-processing method to reduce the difficulty of solving Eq. (\ref{eq1}). Next, we propose a two-step optimization approach for our \textit{discrete flattening} method, where we first solve a coarse optimization problem and then locally refine the obtained solution.

\section{Discrete Flattening} \label{sec_flatten}

To achieve a suitable 2D layout for all triangles where the gap values between adjacent triangles fall within the required range of the shrinkable length, and thus meet the requirement of eliminating precise control over the material’s thermal shrinkage, we introduce our discrete flattening method in this section. This method comprises two steps: coarse optimization and local refinement.

\subsection{Coarse Optimization}

We propose the coarse optimization to solution to the problem in a cost-efficient way. We employ a dynamic adjustment strategy to meet these energy terms, where only arguments and weights are changed during the optimization process without adding other energy terms, which reduces the difficulty of solving our comprehensive optimization goal. Additionally, we introduce some supportive strategies to enhance the solution process. Specifically, we introduce an \textit{auto-cutting} strategy to reduce shape distortion at local positions in the flattened mesh. Moreover, we introduce an \textit{alignment} strategy to reduce intersection of triangles in $M$.

\subsubsection{Auto-Cutting} \label{sec_autocut}

For non-developable surfaces, conformal mapping introduces local distortions, leading to excessive gaps between adjacent triangles (see Fig. \ref{cutting}, left). Therefore, we introduce the \textit{auto-cutting} strategy to address this issue. This strategy involves cutting $M$ without dividing it into multiple patches. Specifically, it entails removing some linkages with excessive gap values during each iteration of optimization while ensuring that $\mathcal{G}$ remains connected. This operation effectively releases tension caused by shape distortion of the conformal mapping. Therefore, the gap values of the retained linkages can be better optimized towards the target value. The specific process is as follows:
\begin{itemize}
    \item Mark existed linkages $L$ as \textit{retained}.
    \item Compute gap values $\mathbf{A}$ of existed $L$ after $i^{th}$ optimization, where the gap value $\mathbf{a}(l) \in \mathbf{A}$.
    \item If $\mathbf{a}(l) > \epsilon_{len}$, mark $l$ as \textit{selected}.
    \item Remove all $l(\textit{selected})$ from $L$ iteratively, If $\mathcal{G}$ is not connected, retain the $l(\textit{selected})$.
\end{itemize}
In the process, The threshold $\epsilon_{len}$ is used to select linkages with the excessive gap values as follows:
\begin{equation}
    \epsilon_{len} = \mathrm{Max} \left(\left( 1 + \epsilon_{tor} \right) \cdot \mathbf{d}, \left( 1 - \mathbf{c} \right) \cdot \mathbf{a_{max}}  \right)
\end{equation}
where $\epsilon_{tor}$ denotes the tolerance of optimization and $\mathbf{c} \in [0, 1]$ is the cutting rate. We use $\mathbf{a_{max}} = \mathrm{Max}(\mathbf{a}(L(\textit{selected})))$ to represent the maximum gap value among linkages, where $l(\textit{selected}) \in L(\textit{selected})$.

To effectively determine the connectivity of $\mathcal{G}$, we employ the Depth-First Search (DFS) algorithm to visit nodes. The main process is as follows:
\begin{itemize}
    \item Construct a graph $\mathcal{G}: (\mathcal{N}, \mathcal{L})$, where $\mathcal{L}$ is formed based on retained $L$. Mark all nodes as \textit{unvisited}.
    \item Using a recursive algorithm, the process begins from a randomly selected node, visiting the next connected node and marking it as \textit{visited}, continuing until all reachable nodes are searched.
    \item If $N_{\mathcal{N}(\textit{visited})} = N_{T}$, $\mathcal{G}$ is connected, else, is not connected.
\end{itemize}

Figure \ref{cutting} illustrates the results with different cutting rates. When the cutting rate $\mathbf{c}$ is low, the flattened result tends to have larger gap values, whereas a higher $\mathbf{c}$ results in smaller gap values (see Fig. \ref{cutting}). Therefore, $\mathbf{c}$ can be set based on the maximum gap value of the result and the stability of the flat plate during assembly, ensuring that any two non-adjacent nodes in $\mathcal{G}$ are connected by at least two paths, as long as the target gap is satisfied as possible.

\subsubsection{Alignment} \label{sec_align}

\begin{figure}[ht]
    \centering\includegraphics[width=\linewidth]{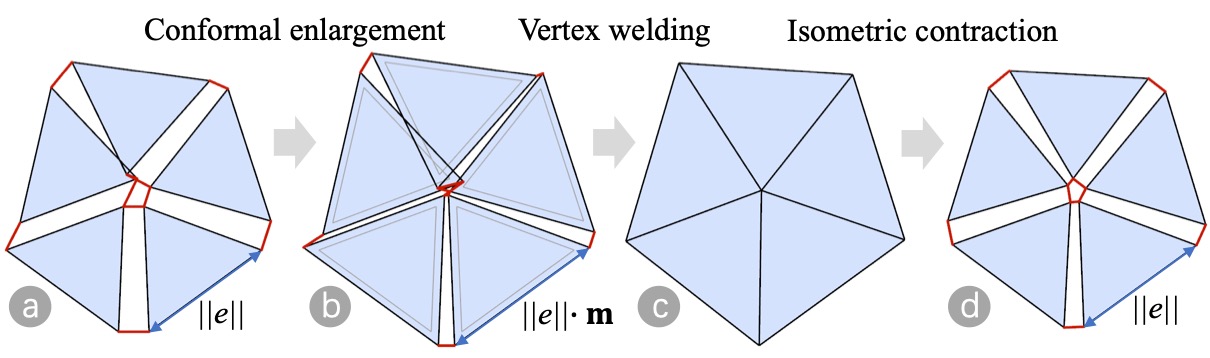}
    \caption{Core idea of \textit{alignment}: intersection avoidance by magnifying triangles and welding vertices.}
    \label{alignment}
\end{figure}

To address the issue of intersections between triangles, we propose the \textit{alignment} strategy. This strategy involves dynamically adjusting the arguments and weights of the energy terms during the optimization process to effectively guide the solution.
Figure \ref{alignment} shows the core idea of our strategy. First, we enlarge the target edge lengths when solving Eq. \ref{eq2} to increase the size of the triangles (see Fig. \ref{alignment}b). Next, we weld the vertices connected by retained linkages together to eliminate intersections (see Fig. \ref{alignment}c). As a result, $M$ will contain gaps between the adjacent original triangles. Consequently, when we optimize the triangles to their original target edge lengths, namely through isometric contraction, the adjacent triangles retain gaps without intersecting (see Fig. \ref{alignment}d).

Therefore, We set the amplification factor $\gamma = \mathbf{m}$ in $E_{\mathrm{rigid}}$ to enlarge triangles and set the scaling factor $\theta = 0$ in $E_{\mathrm{gap}}$ to pull adjacent triangles as close together as possible. Since the vertices welding, the adjacent edges are aligned to together, thus, we can set the weight $\omega_{\mathrm{fair}} = 0$ in $E_{\mathrm{opti}}$ to eliminate the effect of this energy term (this term can be removed in the program to reduce computational cost). The main process is as follows:
\begin{itemize}
    \item Solve $E_{\mathrm{opti}}$, where $\gamma = \mathbf{m}, \theta = 0, \omega_{\mathrm{fair}} = 0$.
    \item Compute existed maximum gap value $\mathbf{a_{max}}$.
    \item \textit{Auto-Cutting} (Section \ref{sec_autocut}).
    \item Weld vertices.
    \item If $\mathbf{a_{max}} > \epsilon_{tor}$, repeat the process.
\end{itemize}

The amplification factor $\gamma = \mathbf{m}$ for the target edge lengths affects the gap values in subsequent isometric contraction of the triangles. Therefore, $m$ must satisfy the relationship as follows:
\begin{equation}
    \mathbf{m} = 1 + \sqrt{3} \cdot \mathbf{d} / \mathbf{\bar{E}}
    \label{eq7}
\end{equation}
where the function related to $\mathbf{m}$ and $\mathbf{d}$ satisfies the properties of the equilateral triangle. In subsequent isometric contraction, the gap value can be adjusted towards the target value $\mathbf{d}$.

\begin{figure}[ht]
    \centering
    \includegraphics[width=\linewidth]{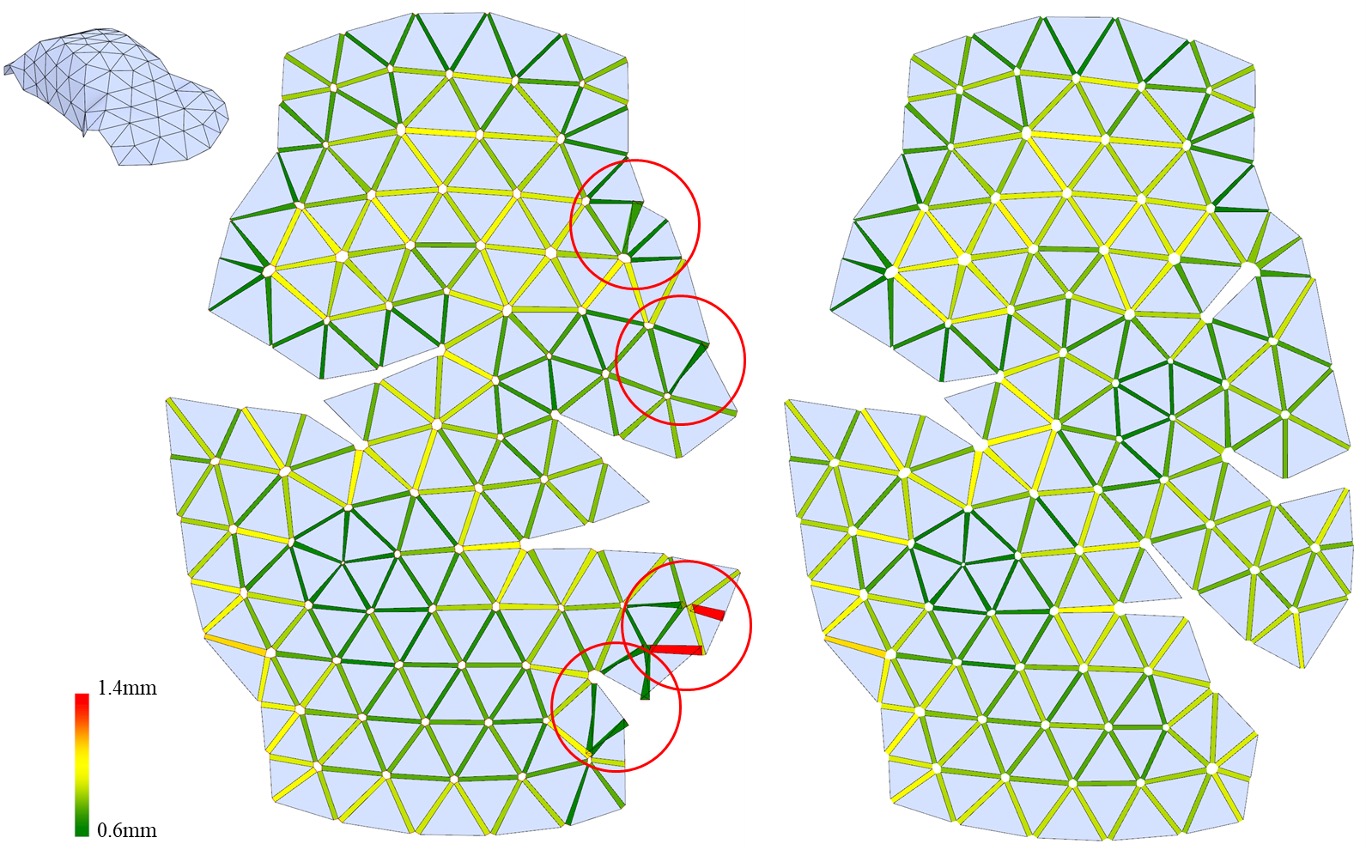}
    \caption{Comparison of results without (left) and with (right) the alignment strategy, where $\mathbf{c} = 0.1$ and the intersection issue is shown within the circle.}
    \label{alignment2}
    \vspace{-15pt}
\end{figure}

After the \textit{alignment} process, we solve Eq. (\ref{eq1}) to isometric contraction, where these arguments are updated: $\gamma = 1, \theta = 1$. Figure \ref{alignment2} displays the results both without and with the \textit{alignment} strategy.

\begin{figure}[ht]
    \centering
    \includegraphics[width=\linewidth]{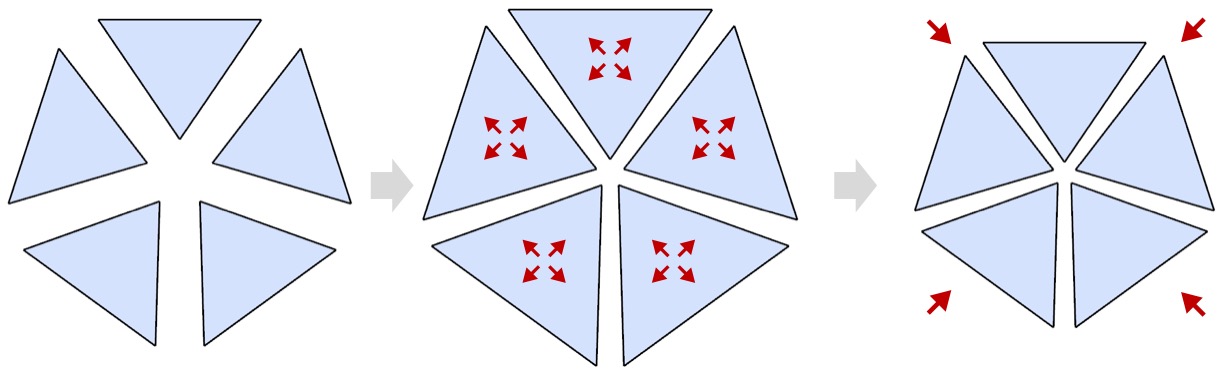}
    \caption{Core idea of the \textit{local refinement}. Starting with the basic 2D layout generated by \textit{coarse optimization} as input (left), we scale each triangle to reduce the gap value (middle). Then, we scale the entire 2D mesh to restore original triangle size (right).}
    \label{refine}
\end{figure}
\subsection{Local Refinement} \label{sec_refine}

The results obtained after solving the coarse optimization as the approximate model meet the desired requirements, but due to the conflict of these energy terms, intersection issues may still exist. We could resolve them by adding an intersecting penalty energy term in Eq. \ref{eq1}, but it would generate performance issues due to the computational cost.
To address this, we employ a \textit{global-to-local} idea. We observe that during the \textit{coarse optimization} phase, the initial iterations establish the basic 2D layout of the triangles, while subsequent iterations focus on optimizing the gaps between adjacent triangles. Therefore, we can significantly reduce computational costs by relaxing the target gap value during the \textit{coarse optimization} to obtain the basic 2D layout and then performing a \textit{local refinement} to optimize the gaps. In this approach, the \textit{coarse optimization} with a relaxed target gap value ensures that the triangles do not intersect. The relaxed target gap value can be achieved by setting a larger target value or by relaxing the termination conditions of the \textit{coarse optimization}.

Specifically, our \textit{local refinement} further optimizes obtained basis 2D layout of triangles by a two-step scaling strategy. Figure \ref{refine} illustrates the core idea. First, we sequentially scale each triangle to reduce the gap value. Then, we scale the entire 2D mesh to obtain the result that restores original triangle size. This operation adjusts only the gap values without altering the sizes of the triangles, helping to achieve results that more closely approximate the target values.

Therefore, we denote a scaling factor as $\lambda$ to scale the each triangle and the 2D mesh. The process of the two-step scaling is as follows:
\begin{itemize}
    \item Scale triangles: $X_{i} = (X_{i} - X_{c}) \cdot \lambda + X_{c}$.
    \item Scale 2D mesh: $X_{i} = (X_{i} - O) \cdot (1/\lambda) + O$.
\end{itemize}
where, each triangle is scaled based on its center point $X_{c}$, that note $X_{c} = (X_{i} + X_{j} + X_{k}) / 3$, $(i, j , k) \in t$. The 2D mesh is scaled based on the world coordinate origin $O: (0,0,0)$.
Therefore, the problem is transformed into a fitting problem. Our goal is to adjust $\lambda$ to make the gaps between triangles fit the target value. We employ gradient descent based on adaptive learning rate to minimize the difference between the average gap value $\mathbf{\bar{A}}$ and the target gap value $\mathbf{d_{t}}$. The process is as follows:
\begin{itemize}
    \item Compute loss: $\epsilon_{loss} = \mathbf{\bar{A}} - \mathbf{d_{t}}$.
    \item Update scale factor: $\lambda = \epsilon_{loss} \cdot \sqrt{\epsilon_{loss}^{2}} \cdot \mathbf{r}_{learn} + 1$.
    \item Two-step scaling.
    \item Repeat until the termination condition is reached.
\end{itemize}
where $\mathbf{r}_{learn}$ is the learning step and $\sqrt{\epsilon_{loss}^{2}}$ indicates the adaptive learning rate.

\begin{figure*}[!t]
    \centering
    \includegraphics[width=\textwidth]{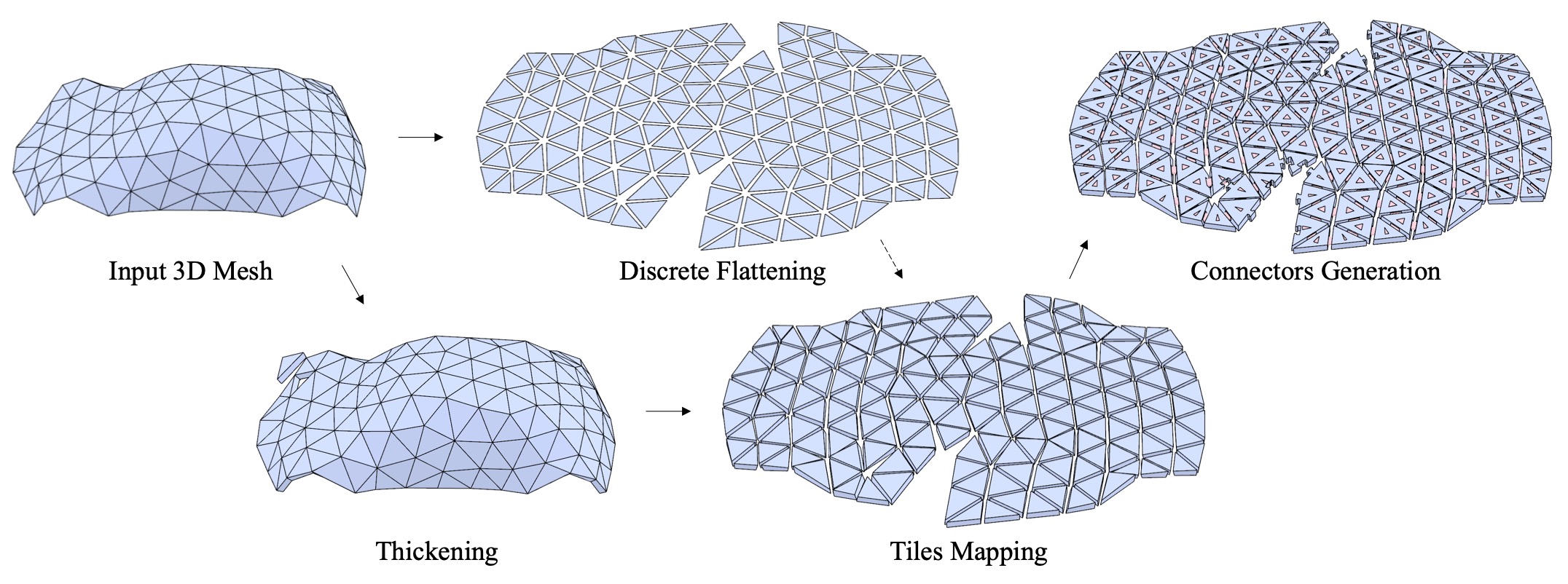}
    \caption{Computational design process of the flat plate. We input the target 3D mesh and the flat plate including tiles and connectors is constructed.}
    \label{structure}
\end{figure*}

Our algorithmic workflow of the \textit{discrete flattening} is shown in Algorithm \ref{alg1}. We set the weights in Eq. (\ref{eq1}) to $\omega_{\mathrm{rigid}} = 100$ and $\omega_{\mathrm{gap}} = \omega_{\mathrm{fair}} = \mathrm{Min}(100, 10+10 \cdot N_{iter})$, where $N_{iter}$ represents the number of iterations. The dynamic weight setting can effectively guide the optimization process. To solve Eq. \ref{eq1}, we use the L-BFGS-B method of the \textit{Scipy} optimization library.
\begin{algorithm}
\caption{Discrete Flattening for 4D printing}
\label{alg1}
\begin{algorithmic}[1]
        \Require{Target 3D mesh $\hat{M}: (\hat{X}, \hat{T})$;}
        \Ensure{Flattened 2D mesh $M: (X, T)$;}
        
        \State $\mathbf{a_{max}} \gets \infty$;
        \State $M \gets$ ConformalMapping($\hat{M}$);
        \State $M \gets$ Alignment($M$); \Comment{Section \ref{sec_align}}
        
        \While{$\mathbf{a_{max}} > (1 + \epsilon_{tor}) \cdot \mathbf{d}$}
            \State Solve Eq. \ref{eq1};
            \State $\mathbf{a_{max}} \gets$ Max($\mathbf{a}(L)$);
            \State $L \gets$ AutoCutting($L$); \Comment{Section \ref{sec_autocut}}
        \EndWhile
        
        \State $M \gets$ LocalRefinement($M, \mathbf{d_{t}}$); \Comment{Section \ref{sec_refine}}
    \end{algorithmic}
\end{algorithm}

\section{Structure Design} \label{sec_structure}

In this section, we explain our structure design details, including flat plate design and parameter selection.

\subsection{Flat Plate Design}

Our goal is to create a suitable flat plate based on the flattened 2D mesh. The flat plate consists of tiles and thermoshrinkable connectors. We use computational design tools to create the flat plate. As illustrated in Figure~\ref{structure}, the process begins with thickening the target 3D mesh to create the tiles. Then, we flatten the target 3D mesh into a 2D mesh using our \textit{discrete flattening} method. The tiles are subsequently mapped onto this 2D mesh to produce a 2D layout for all tiles. Finally, connectors are designed based on the retained linkages (Section~\ref{sec_autocut}).

\begin{figure}[t]
    \centering\includegraphics[width=\linewidth]{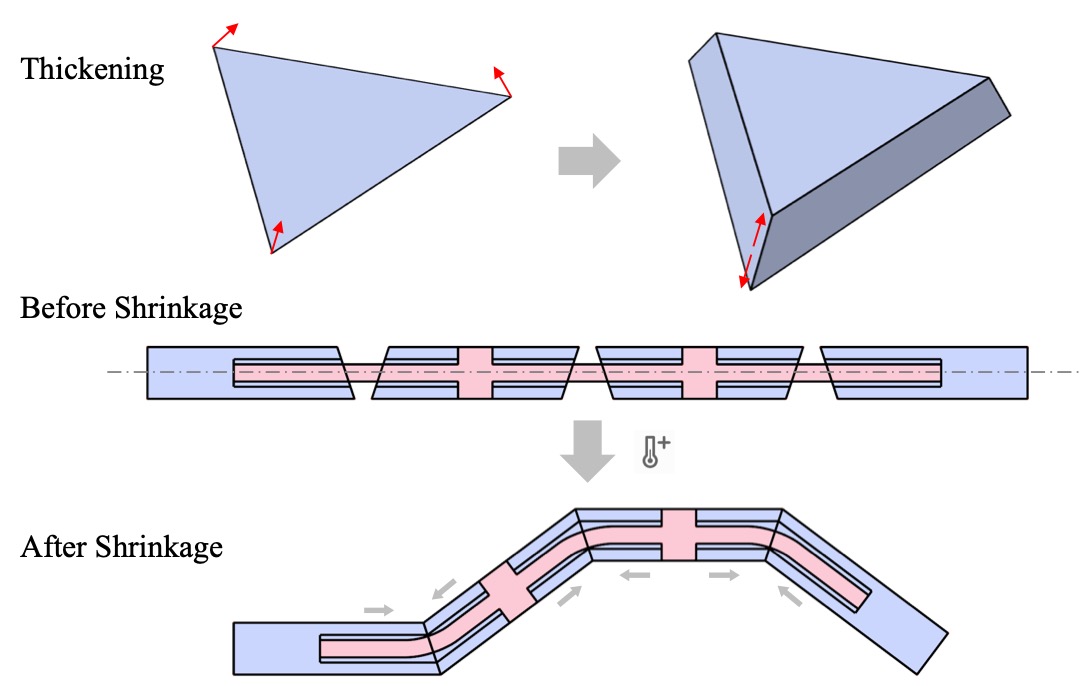}
    \caption{Tiles are formed by Thickening triangles on both sides along the normal directions, which induces bending in different directions.}
    \label{mid}
\end{figure}

\paragraph{Tiles Generation}
We thicken each triangle in the target 3D mesh on both sides along the directions of vertex normals, incorporating connectors in the middle of the tiles. As illustrated in Figure~\ref{mid}, this design allows the flat plate to bend on both sides as the connectors contract. Subsequently, these tiles are mapped onto the flattened mesh form the target 3D mesh.

\begin{figure}[t]
    \centering
    \includegraphics[width=\linewidth]{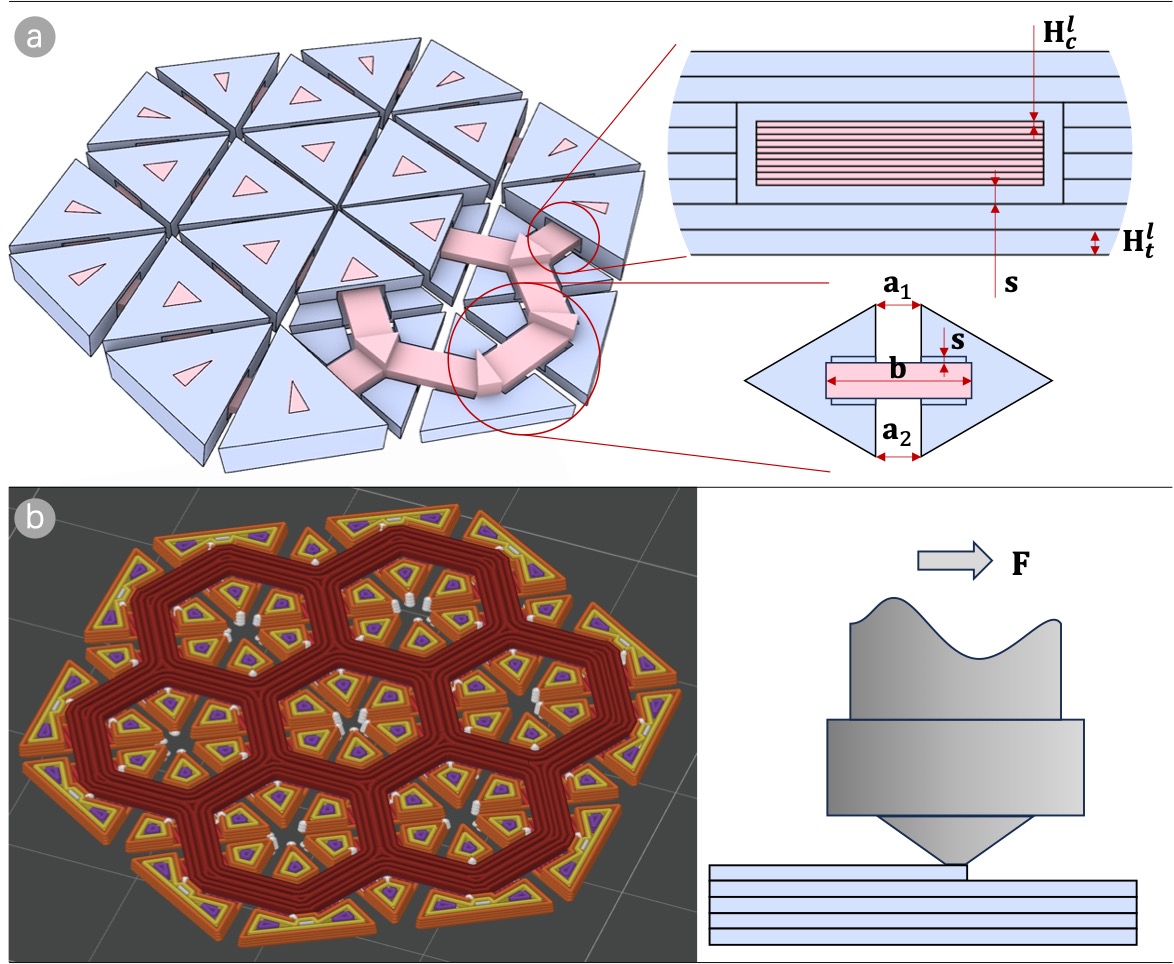}
    \caption{structure of the flat plate. (a) Cross-sectional structures and parameters of the flat plate. (b) Printing Paths of the flat plate.}
    \label{connector}
\end{figure}

\paragraph{Connectors Generation}
Connectors are used to link adjacent tiles and pull them tightly together. We design these connectors based on retained linkages. As shown in Figure \ref{connector}a, each connector is fixed at the centers of the adjacent triangles. To ensure that the connector's shrinkage capability is not affected, we set space around them within the tiles. The printing paths of the connectors are aligned with the shrinkage direction. We utilize universal slicing software to define these printing paths, which are configured as filled paths using \textit{Concentric} paths (see Fig.~\ref{connector}b). Additionally, we design interlocking structures in adjacent tiles that lack connectors, enabling them to join together after deformation to enhance structural stability. We employ the computational tool \textit{Rhino 3D \& Grasshopper}, a visual node-programming tool, to create the flat plate as a digital model. We write an program within this tool to meet these structural requirements. By inputting the flattened 2D mesh into this program, it automatically generates a digital model of the flat plate.

\subsection{Parameter Selection} \label{sec_param}

The purpose of this section is to obtain suitable structural parameters and printing parameters for the designed flat plate through shrinkage experiments.

\paragraph{Basic parameters}
Figure \ref{connector} illustrates the basic parameters of the flat plate. For structural parameters, the connector length is denoted as $\mathbf{b}$, and the gap value $\mathbf{a}$ of the adjacent tiles is measured as the average of $\mathbf{a}{1}$ and $\mathbf{a}{2}$. We define a gap rate as $\mathbf{r} = \mathbf{a} / \mathbf{b}$, which is related to the shrinkage rate. After heating, we denote the gap after contraction as $\mathbf{a}^{'}$. Additionally, the space between the tile and the connector is denoted as $\mathbf{s}$ to prevent the connector from sticking to the tile during the printing process.
For printing parameters, the printing layer thickness of the connector and the tile are denoted as $\mathbf{H}_{c}^{l}$ and $\mathbf{H}_{t}^{l}$, respectively. The printing speed of the connector is denoted as $\mathbf{F}_{c}$, and the printing speed of the tiles as $\mathbf{F}_{t}$.
We use basic PLA materials to print the flat plate, as it is inexpensive, environmentally friendly. During all experiments, the water temperature is kept constant at 75\textdegree C, and the material is heated for more than 1 minute to ensure complete shrinkage. It is important to note that when the material is in a glassy state, higher water temperatures (e.g., 90°C) result in greater shrinkage and more accurate deformation. We set the water temperature to 75\textdegree C as the lower limit to enhance robustness of our method.

\paragraph{Gap between Adjacent Tiles}
To ensure that the connectors can tightly pull adjacent tiles together, we need to determine the appropriate gap value through experimentation. This requires the gap value to be less than the shrinkage length of the material. Therefore, we need to establish the relationship between the gap value and the shrinkage rate.
Specifically, we change $\mathbf{r}$ and control other parameters including $\mathbf{H}_{c}^{l}$ and $\mathbf{s}$. Therefore, we create three groups of samples with different parameters: first group as $G_{1}$ with parameters $\mathbf{H}_{c}^{l} = 0.06$ mm and $\mathbf{s} = 0.36$ mm; second group as $G_{2}$ with parameters $\mathbf{H}_{c}^{l} = 0.06$ mm and $\mathbf{s} = 0.3$ mm and third group as $G_{3}$ with parameters $\mathbf{H}_{c}^{l} = 0.08$ mm and $\mathbf{s} = 0.3$ mm. Based on previous research~\cite{vanmanen2017, rajkumar2018, kacergis2019}, the faster the printing speed, the greater the material's shrinkage capacity. Therefore, within the printing speeds supported by most printers, we set the printing speed for the connectors to $\mathbf{F}_{c} = 100$ mm/s.
\begin{figure}[t]
    \centering
    \includegraphics[width=\linewidth]{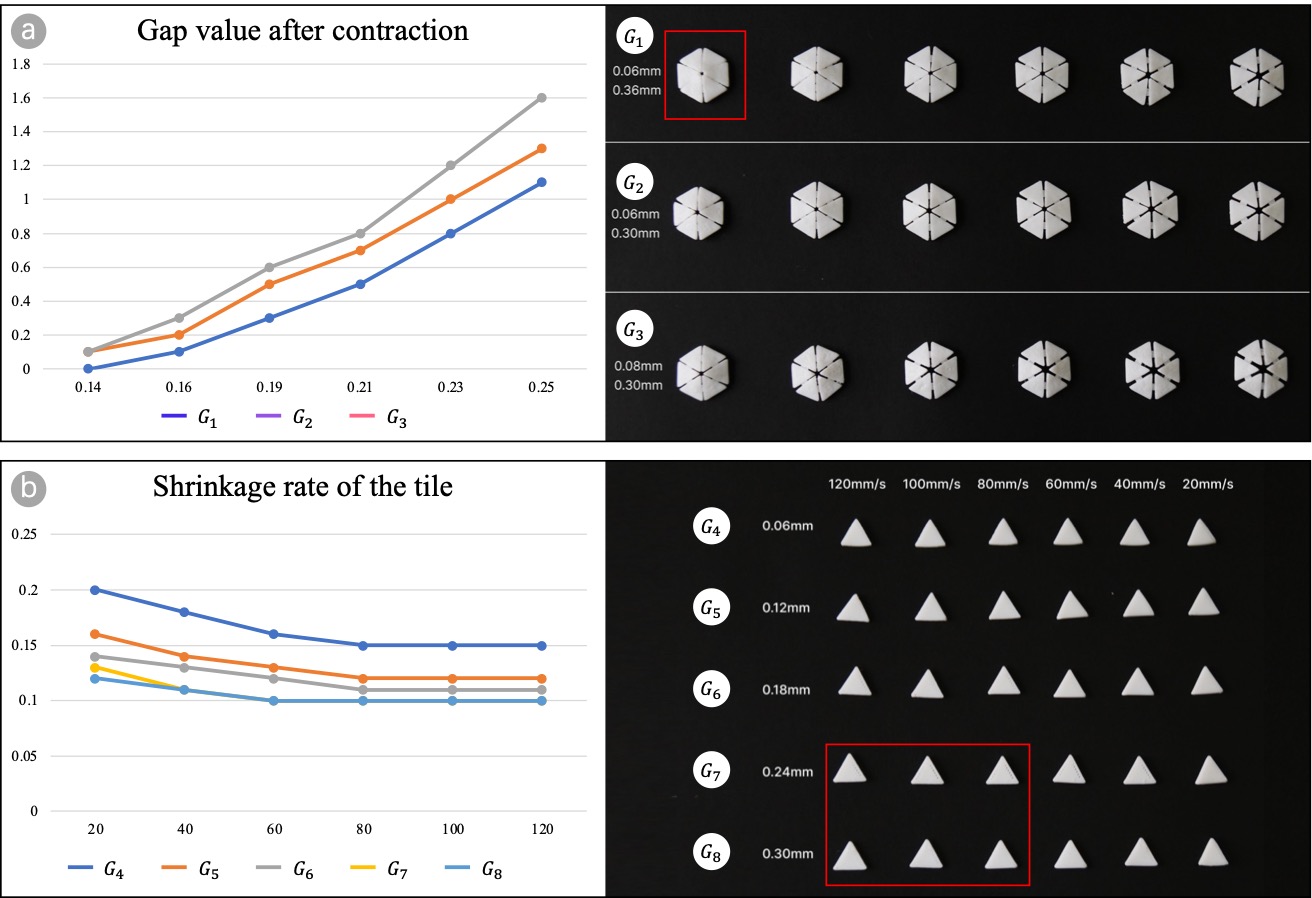}
    \caption{Parameter selection. (a) Gap values of the flat plate unit with different $\mathbf{r}$ after shrinkage, where $\mathbf{a}^{'} = 0$ is represented as the best result in red rectangle. (b) Shrinkage rate of the tile with different printing speeds, where the results in red rectangle have the smallest shrinkage.}
    \label{param}
    \vspace{-5pt}
\end{figure}

As illustrated in figure \ref{param}a, after heating the flat plate unit in 75\textdegree C water for a minute, $G_{1}$ is the best with smaller $\mathbf{a}^{'}$ value compared to the other groups, and when $\mathbf{r} \le 0.14$, $\mathbf{a}^{'} = 0$, meaning that adjacent tiles are pulled tightly together by the connectors. Note that $\mathbf{s}$ in $G_{1}$ and $G_{2}$ has different values. A smaller $\mathbf{s}$ may cause adhesion between the connectors and tiles during the printing process, preventing the connectors from fully contracting. Therefore, we set parameters of $\mathbf{H}_{c}^{l} = 0.06$ mm and $\mathbf{s} \ge 0.36$ mm as our fixed parameters. The target gap value meets the relationship as follows:
\begin{equation}
    \mathbf{a} = \frac{\mathbf{\bar{E}} \cdot \mathbf{r}}{\sqrt{3} \cdot (1 - \mathbf{r})} 
\end{equation}
where $\mathbf{r} = 0.14$ is represented as the lower limit of the material shrinkage rate and $\mathbf{d}_{t} = \mathbf{a}$. As long as the shrinkage rate of the material is greater than $\mathbf{r}$, the adjacent tiles can fully contact together. The shrinkage of PLA materials tested in previous studies~\cite{wang_4dmesh_2018, gu_geodesy_2019, wang_morphingcircuit_2020, jourdan_shrink_2023} is greater than this value, so these materials can also be used in our method.

\paragraph{Printing Speed of Tiles}
The 3D printing time of the flat plate is primarily constrained by the printing speed of tiles. To reduce fabrication time and maintain the tile shape after heating, we determine the appropriate printing speed for tiles through experimentation. Specifically, we change $\mathbf{F}_{t}$ of the tile and control parameter of $\mathbf{H}_{t}^{l}$. Therefore, we create five groups of samples with different $\mathbf{F}_{t}$ (from $G_{4}$ to $G_{8}$), and $\mathbf{H}_{t}^{l}$ in each groups is increasing.

After heating the flat plate unit, we denote the edge length after heating as $||{e}^{'}||$. As illustrated in figure \ref{param}b, $G_{8}$ with parameters of $\mathbf{H}_{t}^{l} = 0.3$ mm and $\mathbf{F}_{tile} \ge 80$ mm/s, has smaller contraction compared to the other groups. Note that when the printing speed exceeds 80 mm/s, the tile no longer shrinks. In comparison to the printing speed, the printing layer thickness has a greater impact on shrinkage. Therefore, when printing tiles, we can choose a thicker layer thickness (e.g., $\mathbf{H}_{t}^{l} \ge 0.24$ mm) and a faster printing speed (e.g., $\mathbf{F}_{t} \ge 100$ mm/s) to improve fabrication efficiency.

\section{Fabrication and Evaluation} \label{sec_implement}

In this section, we explain implementation details, including the fabrication process and evaluation.
\begin{figure*}[!t]
    \centering
    \includegraphics[width=\textwidth]
    {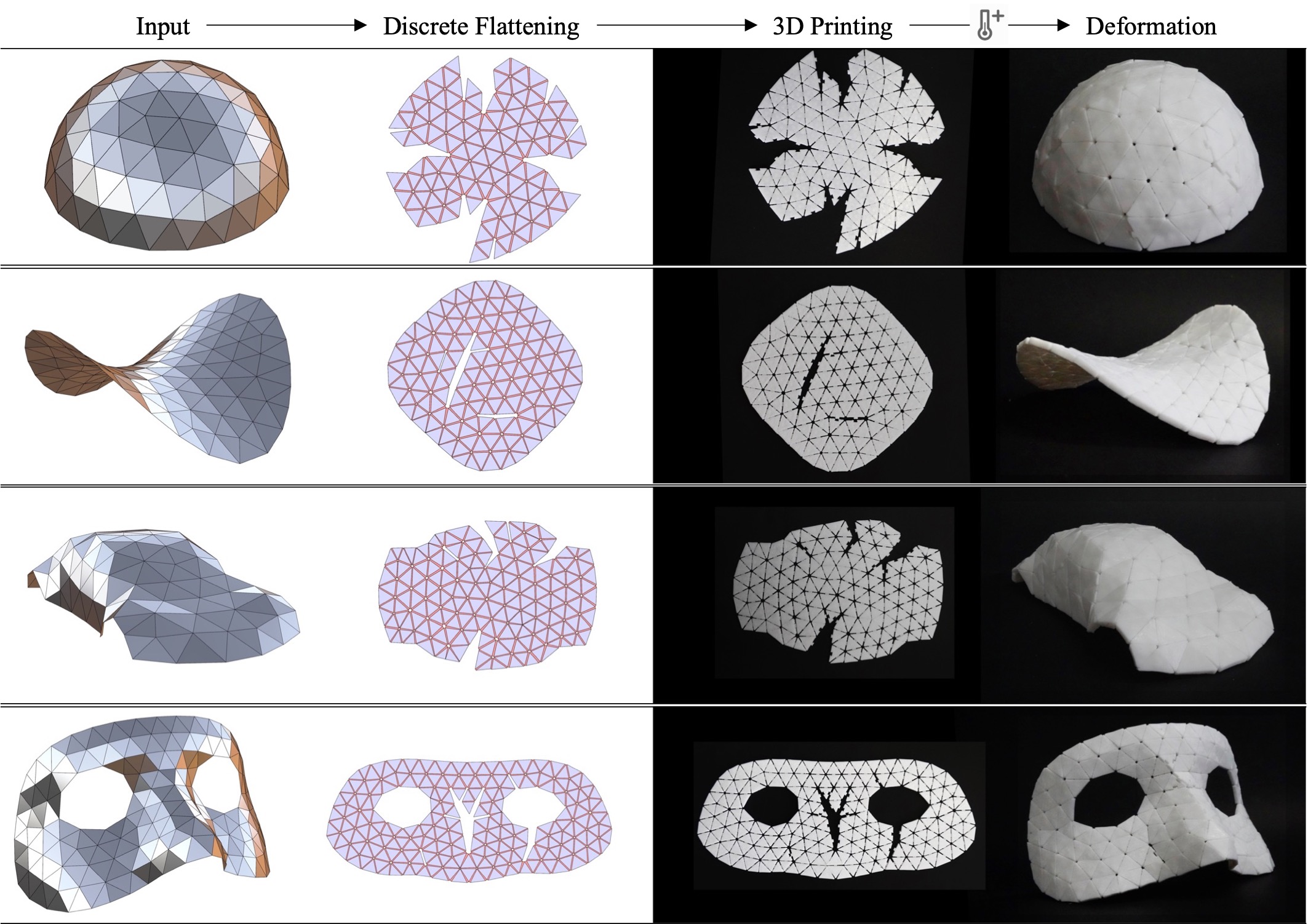}
    \caption{The process of our 4D printing method from design to fabrication. Tested models are \textit{hemisphere}, \textit{hyperboloid}, \textit{car} and \textit{mask}, respectively.}
    \label{result1}
\end{figure*}

\subsection{Fabrication}

Our fabrication process begins with 3D printing the flat plate, including tiles and connectors, using a Bambu Lab X1 3D printer. The flat plate requires no support structures and utilizes a single material. We use the parameters provided in Section \ref{sec_param} for 3D printing.
After printing, we place the flat plate into 85°C constant temperature water. The flat plate gradually self-shapes into the target shell. After the self-shaping process is complete, if the flat plate has any cuts formed by \textit{auto-cutting}, we need to interlock the adjacent tiles without connectors before the material fully hardens.

\begin{figure*}[!t]
    \centering
    \includegraphics[width=\textwidth]{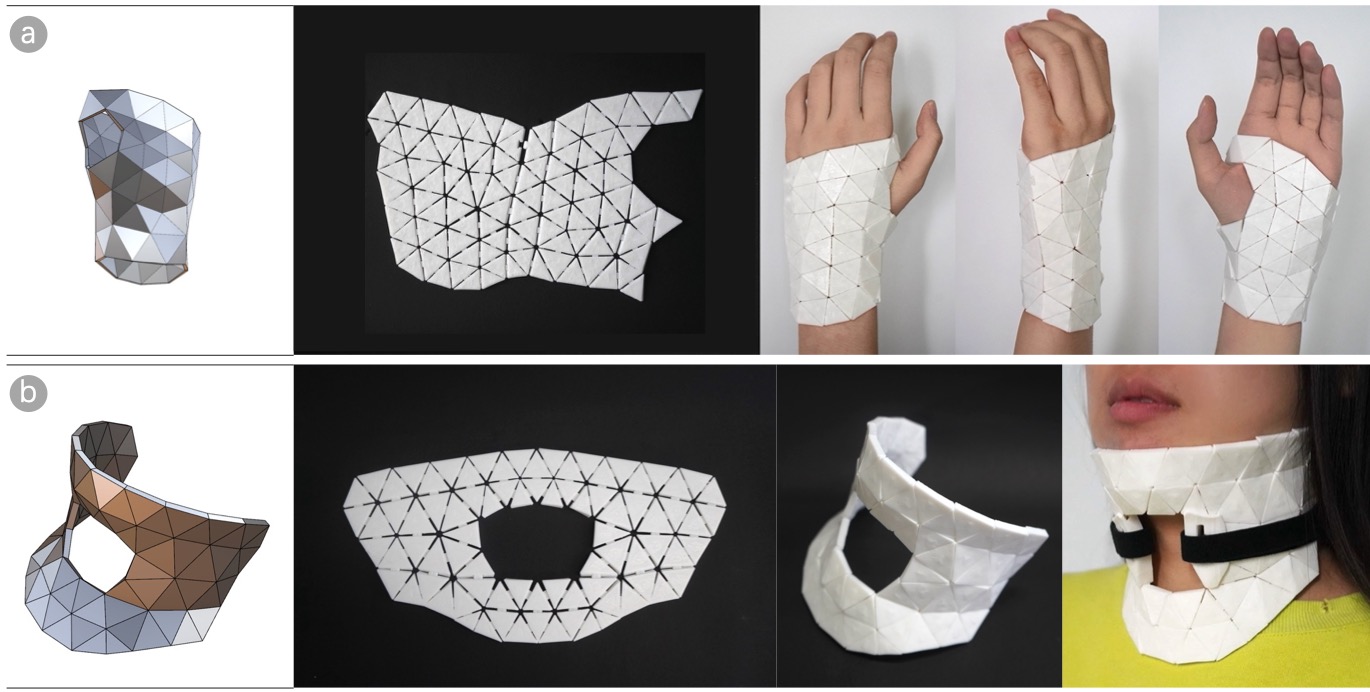}
    \caption{Applications of the medical supports. (a) The wrist brace is used to stabilize the wrist joint. (b) The neck brace is used to support the head.}
    \label{application}
    \vspace{-3mm}
\end{figure*}
\subsection{Evaluation}

\paragraph{Applications} 
We demonstrate the effectiveness of the proposed method through the printing results of various shapes, as shown in Figure~\ref{result1}. The first two rows feature a \textit{hemisphere} and a \textit{hyperboloid}, which have positive and negative Gaussian curvatures, respectively. The third row displays a \textit{car} with a more complex surface and spatially varying curvatures. The last row presents a \textit{mask} designed by a professional, featuring holes and areas of high local curvature, illustrating the applicability of our method to intricate shapes. These examples highlight the versatility of our method for different complex geometries. Furthermore, we demonstrate its use in medical appliances that conform to the body's curvature and help fix patients' joints (Fig.~\ref{application}).

\paragraph{Performance}
We present the performance of our method via computational time and printing time as shown in Table~\ref{tab1}. The computational time typically takes less than two minutes, where the \textit{discrete flattening} is usually the most time-consuming step. Printing usually takes ranges around three hours, depending on the size. Note that both the processes of \textit{discrete flattening} and the \textit{structure design} run autonomously. Once a digital model of the flat plate is generated, we use slicing software to control a 3D printer to fabricate it. Printing time could be further reduced by increasing the printing speed or decreasing the thickness of each layer for the tiles.
\begin{table}[ht]
\centering
\caption{For each model, we show the number of faces, the total thickness of the flat plate, the area in the flat plate, the computational time for computing the flat plate, the required 3D printing time, and the distance between the fabricated shell and the target.}
\resizebox{\linewidth}{!}{%
\begin{tabular}{lcccccc}
\toprule
\textbf{Model} & \textbf{Face} & \textbf{\makecell{thickness \\ ($mm$)}} & \textbf{Area ($cm^{2}$)} & \textbf{\makecell{Computational \\ time ($s$)}} & \textbf{\makecell{Printing \\ time ($min$)}} & \textbf{\makecell{Distance to target \\ (avg // max) ($mm$)}} \\
\midrule
Half-sphere & 143 & 2.4 & 99.83 & 64.05 & 150 & 0.49 // 1.98 \\
Hyperboloid & 162 & 2.4 & 101.57 & 17.02 & 160 & 1.38 // 6.58 \\
Car & 188 & 2.4 & 112.18 & 36.24 & 158 & 0.65 // 4.46 \\
Mask & 241 & 2.4 & 207.90 & 49.36 & 228 & 1.59 // 5.59 \\
\bottomrule
\end{tabular}
}
\label{tab1}
\vspace{-5pt}
\end{table}

\begin{figure}[ht]
    \centering
    \includegraphics[width=\linewidth]{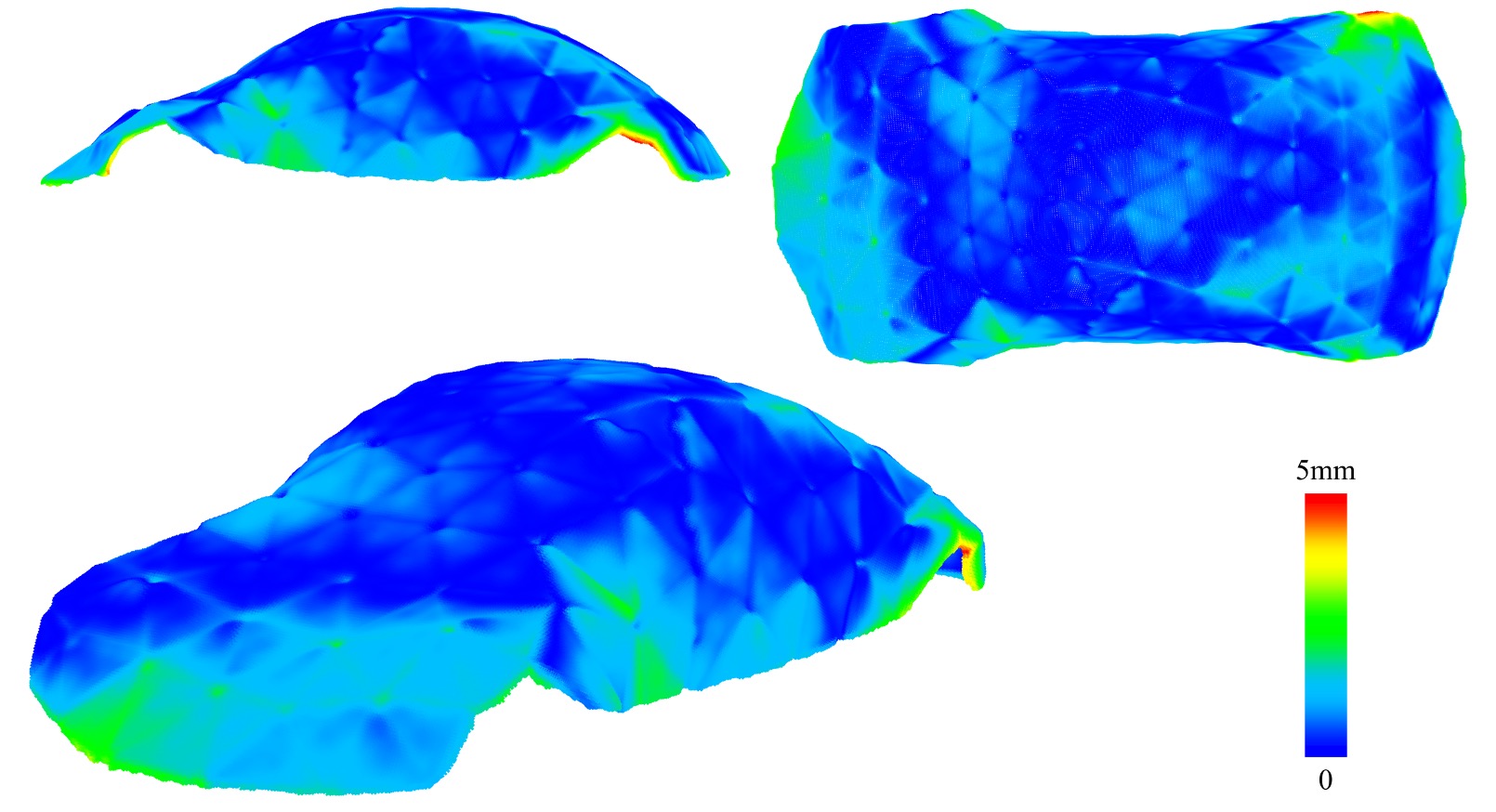}
    \caption{Error visualization of a 3D scan of the \textit{car} model, where the distance error between each point and the target mesh is shown as color-coded from blue to red.}
    \label{vis}
\end{figure}

\begin{figure}[ht]
    \centering
    \includegraphics[width=\linewidth]{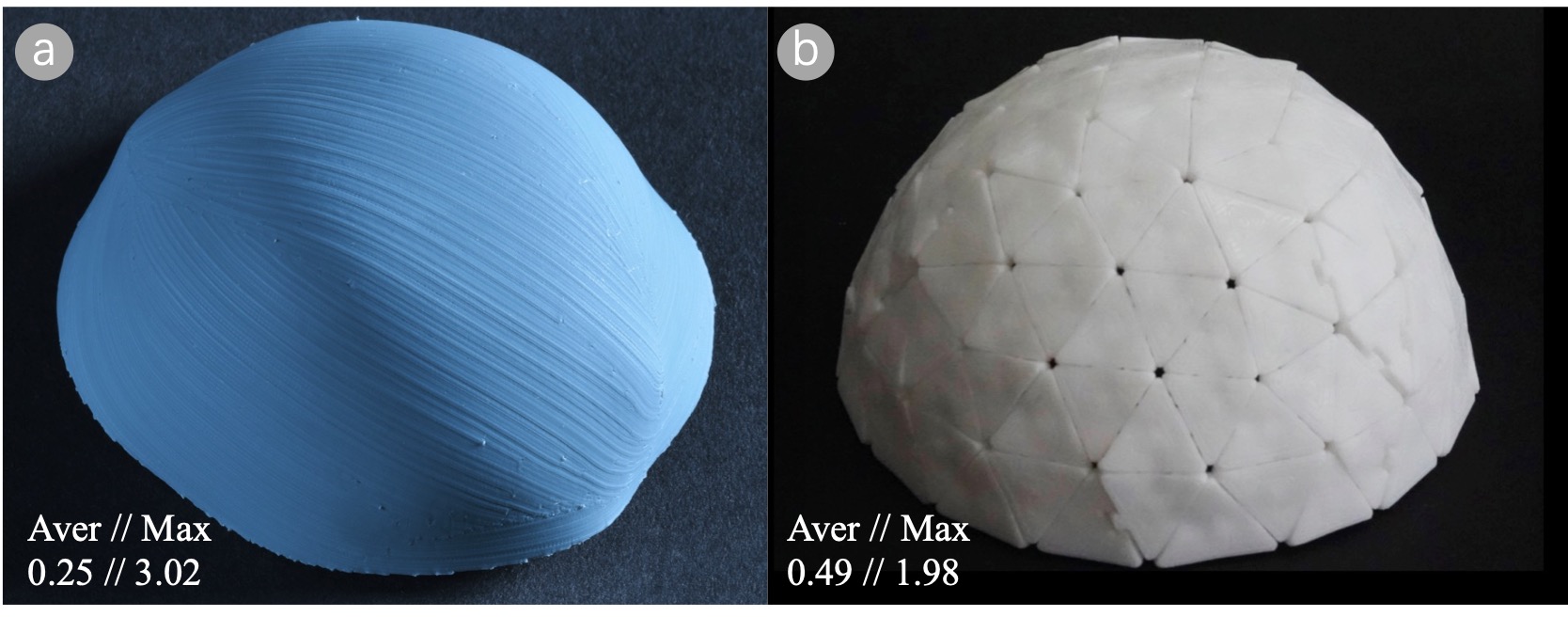}
    \caption{Comparison of two results. Jourdan et al.~\cite{jourdan_shrink_2023} (a) and our method (b).}
    \label{compare}
\end{figure}

\paragraph{Accuracy}
To evaluate accuracy, we compared the fabricated results with the target shapes by estimating the error. Using the \textit{Reopoint POP 3} laser scanner, we scanned and reconstructed the fabricated shells as point clouds. We then measured the distance between each point of the reconstructed model and the target mesh. Table \ref{tab1} presents the distance errors for these shapes. Specifically, for the \textit{car} model, the error distribution is visualized in Figure~\ref{vis}, with colors ranging from blue (no error) to red (errors clamped at 5 mm). The average distance error for the \textit{car} model is below 0.5 mm. We also scanned a complex model, the \textit{mask}, which is approximately 25 cm in size. The average error was below 1.6 mm, with a maximum error of 5.59 mm due to the tendency of the mask's sides to bend. In scenarios requiring a tight fit to another object, the natural flexibility and strength of the shell enable it to snap into place easily.
Addtionally, we also compare our result with those produced by Jourdan et al.~\cite{jourdan_shrink_2023} by fabricating the same size of the \textit{hemisphere}. As illustrated in Figure~\ref{compare}, the average error for both method is within 0.5 mm. However, our result shows a smaller maximum error compared to theirs. Figure~\ref{compare}a shows the inaccurate boundaries of the hemispheres in Jourdan's method. The cause of this phenomenon could be non-uniform shrinkage rate resulting from anisotropic printing paths.

\begin{figure}[ht]
    \centering
    \includegraphics[width=\linewidth]{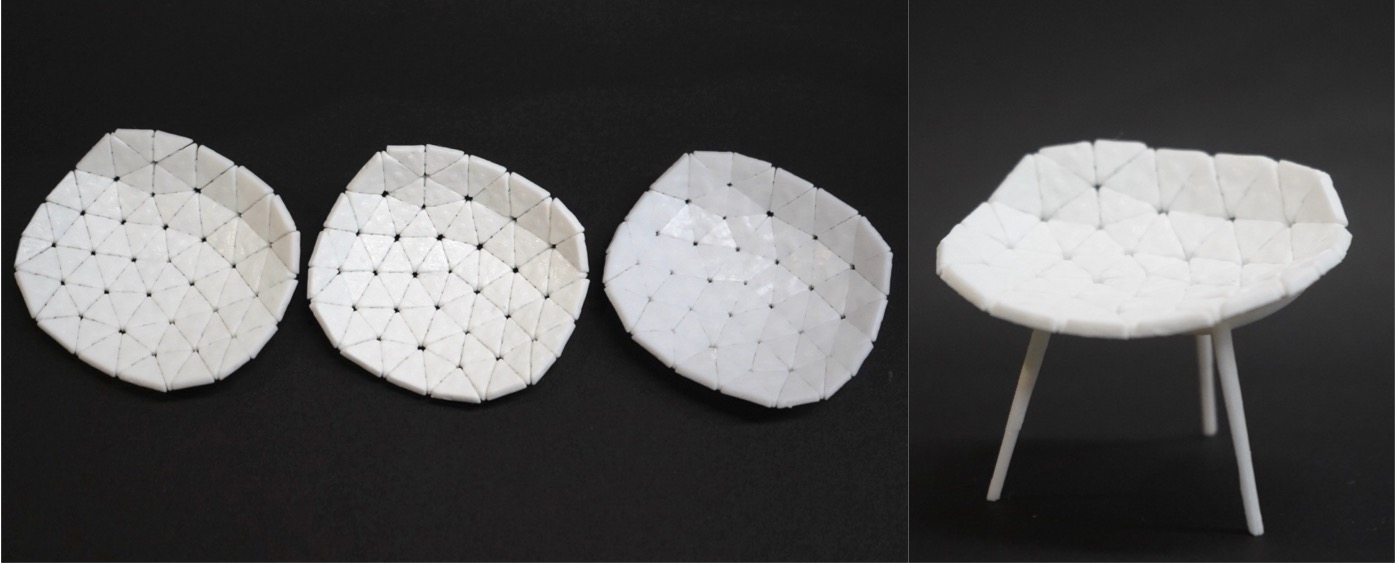}
    \caption{Similar fabricated results with different shrinkage rates, where these average values of the shrinkage rate are not less than 0.14 (0.14, 0.19 and 0.15, respectively). The deformation is completed in hot water at 75\textdegree C.}
    \label{similar}
    \vspace{-10pt}
\end{figure}

\paragraph{Robustness}
To evaluate the robustness of our method with materials exhibiting different shrinkage rates, we used PLA from various vendors. Table \ref{vendors} displays the shrinkage rates of PLA materials from different vendors at various heating temperatures. We fabricated flat sheets with size of 20 mm, a layer thickness of 0.06 mm, and a printing speed of 100 mm/s. After deformation, we calculated the shrinkage rates by measuring the ratio of the length after contraction to the length before contraction. We chose materials with an average shrinkage rate of at least 0.14 to fabricate the chair surfaces. Figure \ref{similar} compares the fabricated results with different shrinkage rates, showing similar outcomes. This indicates that our method does not depend on material properties, as long as the material's shrinkage rate exceeds a certain threshold.
\begin{table}[ht]
\centering
\caption{Comparison of the shrinkage rates of different vendors.}
\resizebox{\linewidth}{!}{%
\begin{tabular}{c c c c c c c}
\toprule
\textbf{Temperature} & \textbf{Bamboo} & \textbf{Polylite} & \textbf{Tinmorry} & \textbf{Polyterra} & \textbf{Esun} & \textbf{Sanlv} \\
\midrule
70\textdegree C & 0.10 & 0.13 & 0.10 & 0.04 & 0.11 & 0.11 \\
75\textdegree C & 0.14 & 0.18 & 0.13 & 0.04 & 0.12 & 0.14 \\
80\textdegree C & 0.18 & 0.27 & 0.14 & 0.06 & 0.13 & 0.21 \\
\midrule
\textbf{Average} & 0.14 & 0.19 & 0.12 & 0.05 & 0.12 & 0.15 \\
\bottomrule
\end{tabular}
}
\label{vendors}
\vspace{-10pt}
\end{table}

\begin{figure}[ht]
    \centering
    \includegraphics[width=\linewidth]{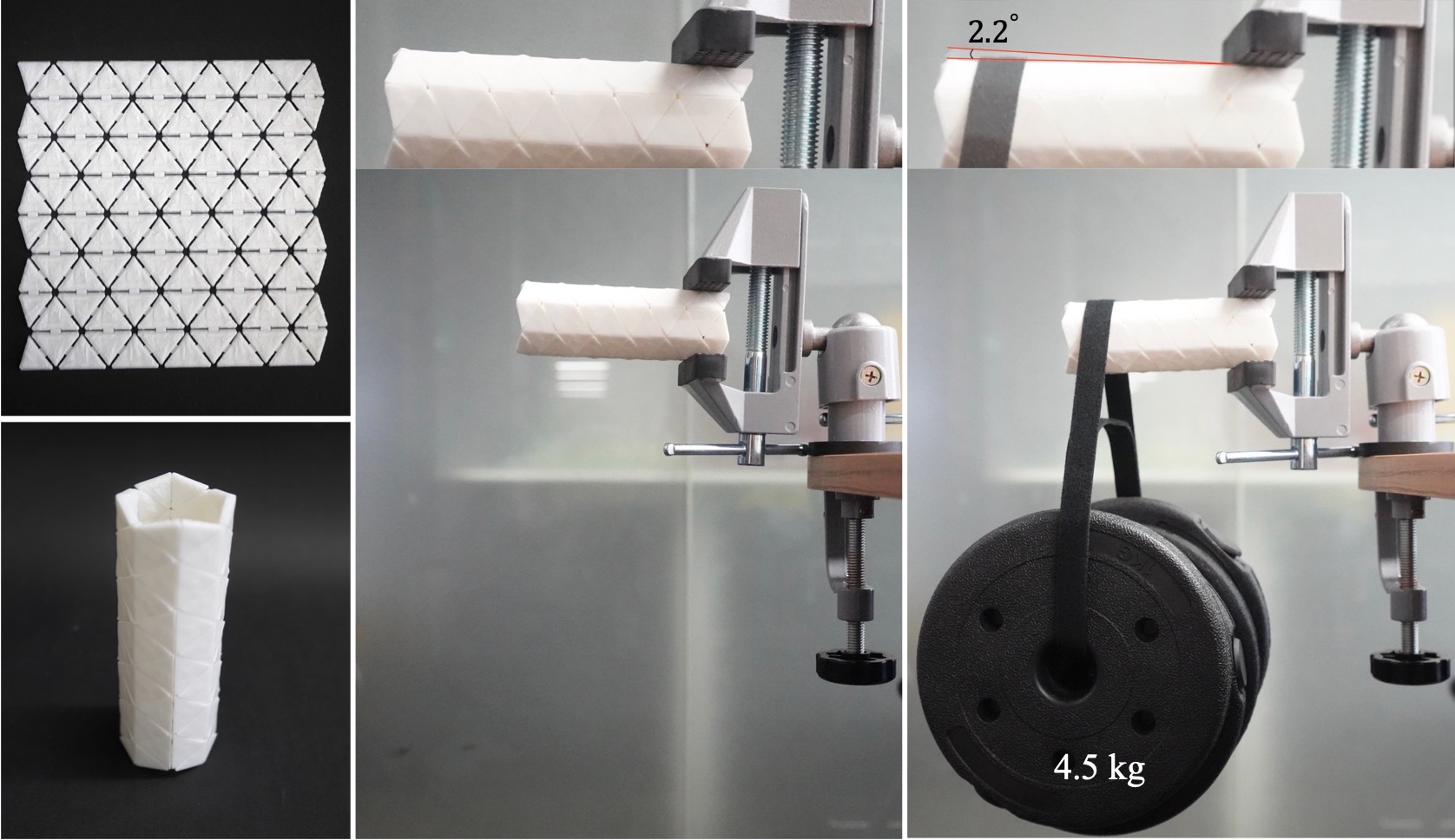}
    \caption{Stiffness test by applying a load to simulate a cantilever bending.}
    \label{stiffness}
\end{figure}

\paragraph{Stiffness}
To test the mechanical stability of the structure created using our method, we evaluated the stiffness by applying a load to simulate a cantilever bending test. We fabricated a pipe with a tile thickness of 3 mm and a connector thickness of 1.2 mm. One end of the pipe was fixed with a clamp, while the other end bore a load of 4.5 kg. The result, shown in Fig.~\ref{stiffness}, indicates that the end of the pipe bent 2.2 degrees relative to the clamping point. Based on this result, our method is suitable for fabricating products that require mechanical stability.

\section{Discussion} \label{sec8}

We have presented FreeShell, a robust 4D printing technique that uses thermoshrinkage to create precise 3D shells. The core idea is to compute a suitable 2D layout for all tiles by our \textit{discrete flattening} method, where the gap between adjacent triangles meets the materials' shrinkage. Subsequently, we generated the flat plate including tiles and connectors based on the 2D layout. The flat plate is fabricated by 3D printing and deformation in the heating water to form the target shell. Finally, we tested fabricated results and demonstrated some applications.
Our method eliminates the need for support structures and segmentation, simplifying the fabrication process. It ensures precise deformation as long as the material meets a shrinkage rate threshold, offering more flexible and robust fabrication conditions, i.e., fully context-free. Additionally, our method applies to various curved shells, including those with holes.

\paragraph{Limitations and Future Work}
Although our method yields promising results, there are still some limitations.
During the optimization process, the energy decreases rapidly in the early iterations but slows down in the later steps to meet the target gap requirement, resulting in many iterations for convergence. It leads to longer optimization time for computing a 2D layout, especially when the target mesh has a large number of triangles.
Additionally, while our method can automatically cut shapes to facilitate distortion-free flattening, user-specified cutting has not yet been developed. In the future, we plan to incorporate interactive cutting features to better meet user needs.
Furthermore, when flattening a hyperboloid with nagtive Guassian curvature, the gap value of adjacent triangles at the center of the flat plate larger than the gap value at the boundary. Significant negative Gaussian curvature often causes the boundary triangles to intersect. Therefore, we need to employ the material with significant shrinkage to set larger the target gap value.
During the deformation process, since some flat plates with cuts have interlocking structures in adjacenct tiles without connectors, these structures cannot be assembled by self-shaping. We still need to manually join them together.

For future work, we will enhance algorithm efficiency and add interaction cutting for users. Then, we plan to develop a plugin for the software Rhino \& Grasshopper to benefit use for general designers. Additionally, we plan to extend the structure possibility of the flat plate, such as tiles with different sizes to improve approximation accuracy of high frequency details, or tiles with high frequency details on one side of the tiles. We also plan to explore the method possibility of combining our method of tile contact with continuous deformation methods, such as contraction and bending based on local shrinkage variation in materials.

\acknowledgments{
This work was supported by NSFC 62072338, 62061136003 and NSF Shanghai 20ZR1461500.}

\bibliographystyle{abbrv-doi-hyperref}

\bibliography{ref}
\end{document}